\documentclass[conference, 10pt]{IEEEtran}
\usepackage{amsmath}
\usepackage{url}
\usepackage{graphicx}
\usepackage{color}
\usepackage{placeins}
\usepackage{float}
\usepackage{tabularx,colortbl}
\usepackage{booktabs}
\usepackage{threeparttable}
\usepackage{setspace}
\usepackage[skip=2pt, font=normalsize]{caption}
\newcommand*{\Scale}[2][4]{\scalebox{#1}{$#2$}}%
\let\labelindent\relax
\usepackage{enumitem}
\usepackage{url}
\usepackage{hyperref}

\begin{document}

\onecolumn % make sure you keep this coverpage as one column. In this template, we force the coverpage to be one column with this command and then switch to double column for the remaining of the paper with the \doublecolumn command. 

\begin{description}[labelindent=1cm,leftmargin=3cm,style=multiline]

\item[\textbf{Citation}]{Z. Wang, D. Temel and G. AlRegib, "Fault detection using color blending and color transformations," 2014 IEEE Global Conference on Signal and Information Processing (GlobalSIP), Atlanta, GA, 2014, pp. 999-1003.} \\

\item[\textbf{DOI}]{\url{https://doi.org/10.1109/GlobalSIP.2014.7032271}} \\

\item[\textbf{Review}]{Date added to IEEE Xplore: February 9, 2015} \\

\item[\textbf{Code/Poster}]{\url{https://ghassanalregib.com/publications/}} \\

\item[\textbf{Bib}] {
@INPROCEEDINGS\{Temel2014\_GlobalSIP\_Seismic,\\ 
author=\{Z. Wang and D. Temel and G. AlRegib\},\\ 
booktitle=\{2014 IEEE Global Conference on Signal and Information Processing (GlobalSIP)\},\\ 
title=\{Fault detection using color blending and color transformations\},\\ 
year=\{2014\},\\
pages=\{999-1003\},\\ 
doi=\{10.1109/GlobalSIP.2014.7032271\},\\
month=\{Dec\},\}\\
} \\

\item[\textbf{Copyright}]{\textcopyright 2014 IEEE. Personal use of this material is permitted. Permission from IEEE must be obtained for all other uses, in any current or future media, including reprinting/republishing this material for advertising or promotional purposes,
creating new collective works, for resale or redistribution to servers or lists, or reuse of any copyrighted component
of this work in other works. } \\

\item[\textbf{Contact}]{\href{mailto:alregib@gatech.edu}{alregib@gatech.edu}~~~~~~~\url{https://ghassanalregib.com/}\\ \href{mailto:dcantemel@gmail.com}{dcantemel@gmail.com}~~~~~~~\url{http://cantemel.com/}}
\end{description} 

\thispagestyle{empty}
\newpage
\clearpage

\twocolumn

%\onecolumn
\title{Fault Detection Using Color Blending and Color Transformations}
%\title{Semi-automatic Fault Detection in Time Sections Using Color Space Conversion and Weighted Skeletonization}

\author{\IEEEauthorblockN{Zhen Wang, Dogancan Temel, and Ghassan AlRegib}
\IEEEauthorblockA{Center for Energy and Geo Processing - CeGP\\
School of Electrical and Computer Engineering\\
Georgia Institute of Technology, Atlanta, GA, 30332-0250, USA\\
\{zwang313, cantemel, alregib\}@gatech.edu\\}
}


%Telephone: (800) 555--1212, Fax: (888) 555--1212}


\maketitle

\begin{abstract}
%\boldmath
In the field of seismic interpretation, univariate data-based maps are commonly used by interpreters, especially for fault detection. In these maps, contrast between target regions and the background is one of the main factors that affect the accuracy of the interpretation. Since univariate data-based maps are not capable of providing a high contrast representation, to overcome this issue, we turn these univariate data-based maps into multivariate data-based representations using color blending. We blend neighboring time sections, frames that are viewed in the time direction of migrated seismic volumes, as if they corresponded to the red, green, and blue channels of a color image. Furthermore, we apply color transformations to extract more reliable structural information. Experimental results show that the proposed method improves the accuracy of fault detection by limiting the average distance between detected fault lines and the ground truth into one pixel.
\end{abstract}


% keywords

% \fi
\IEEEpeerreviewmaketitle

\section{Introduction}
The displacement of fractures in the earth's crust leads to the formation of faults, which are significant geological structures for hydrocarbon exploration. The movement of low permeability rocks along faults may seal porous reservoir rocks in traps and results in the formation of reservoir regions. Therefore, oil and gas exploration require the accurate detection of faults. Conventionally, experienced interpreters can label faults in collected seismic data. However, the manual interpretation of seismic data is very time consuming and labor intensive, especially with the dramatically growing size of recently acquired datasets. Thus, the design and the implementation of automatic or semi-automatic fault detection methods are catching a renewed interest in both industry and academia.

The characterization of seismic structures is well practiced in the literature of several fields including but not limited to seismology and geology. The texture of faults caused by the movement of rocks, different from the uniform texture of horizons, represents discontinuities along horizons. Many seismic attributes such as semblance~\cite{marfurt1999coherency}, variance~\cite{van2000seismic}, curvature~\cite{boe2010seismic}, and gradient amplitude~\cite{aqrawi2011improved}\cite{song2012facet} have been used to measure discontinuities.
%Although these measures provide a reasonable estimate of discontinuities, they are not sufficient to accurately detect faults, especially if only one seismic attribute is involved.
In addition to these basic attribute-based approaches, a number of more complex methods involving image processing techniques have been proposed to semi-automatically detect faults. Cohen et al.~\cite{cohen2006detection} introduced the use of directional filters to enhance the discontinuity cube and proposed a thinning process to extract one-pixel-width fault lines.
%Admasu et al.~\cite{admasu2006autotracking} utilized log-Gabor filters to filter out noise and tracked faults throughout the seismic volume by fitting active contours.
The Hough transform, as a powerful tool to detect lines and curves in images, was first proposed by AlBinHassan and Manfurt~\cite{albinhassan2003fault} to detect fault lines in vertical sections.
Similarly, the authors in~\cite{jacquemin2005fault} applied the cascaded Hough transform to detect fault surfaces in 3D seismic data. To obtain more reliable results, Wang et al.~\cite{wang2014hough} proposed detecting fault features with the Hough transform, removing noisy features under geological constraints, and labeling fault lines by optimally connecting the remaining features. Moreover, by borrowing the idea of motion vectors and utilizing a small number of detected fault lines, Wang et al.~\cite{wang2014tracking} tracked fault lines throughout the seismic volume with high interpretation efficiency.
%In~\cite{Yan2013automatic}, the authors adopted an ant-colony algorithm to implement the automatic tracking of faults in the sections of seismic data. However, the high computational cost of this method greatly reduces the interpretation efficiency.
Recently, Zhang et al.~\cite{zhang2013fault} proposed automatically detecting faults in time sections by adopting a biometric algorithm used for extracting the veins of human fingers. The intuition comes from the structural similarity between faults and capillary veins.

The methods mentioned so far overlook similarities among the neighboring structures of time and seismic sections and focus only on univariate data. However, to facilitate fault detection for interpreters, we need to increase the contrast of attribute maps using multivariate representations. This methodology has been widely used to synthesize images, in which color spaces could provide more details. Low contrast may result from limited color tones. Therefore, to adjust the contrast of different regions and highlight the regions of interests, the authors in~\cite{dao2011value} increased the number of tones in visualization using \texttt{RGB} blending. Similarly, the authors in~\cite{henderson2008delineation} and~\cite{boe2010seismic} proposed color blending methods to enhance the visualization of geological elements. In addition to color blending, color transformations also have the capability of enhancing visualization by separating chroma channels from the intensity channel that contains structural information. The authors in~\cite{laake2013} adopted color transformations to map images from \texttt{RGB} to \texttt{HSV} and sharpened the representations of seismic attributes in the saturation (\texttt{S}) channel.

In this paper, we combine seismic attribute extraction with color blending and color transformations to enhance fault detection accuracy for interpreters. We first derive the semblance maps of neighboring time sections and blend these maps into a single color representation. Then, we transfer the blended maps into different color spaces to obtain channels that contain structural information. We enhance these channels using smoothing and adaptive histogram equalization. Then, by applying thresholds on these enhanced channels, we highlight fault regions in binary images. Finally, we combine these binary images and perform weighted skeletonization to extract one-pixel-width fault lines. The paper is organized as follows. In section~\ref{sec:method}, we explain, in detail, the proposed algorithm. Section~\ref{sec:results} shows experimental results and we conclude our discussions in Section~\ref{sec:conclusion}.

\section{The Proposed Method}
\label{sec:method}
\begin{figure}[t]
\begin{minipage}[b]{1.0\linewidth}
\centering
\centerline{\includegraphics[width=9cm]{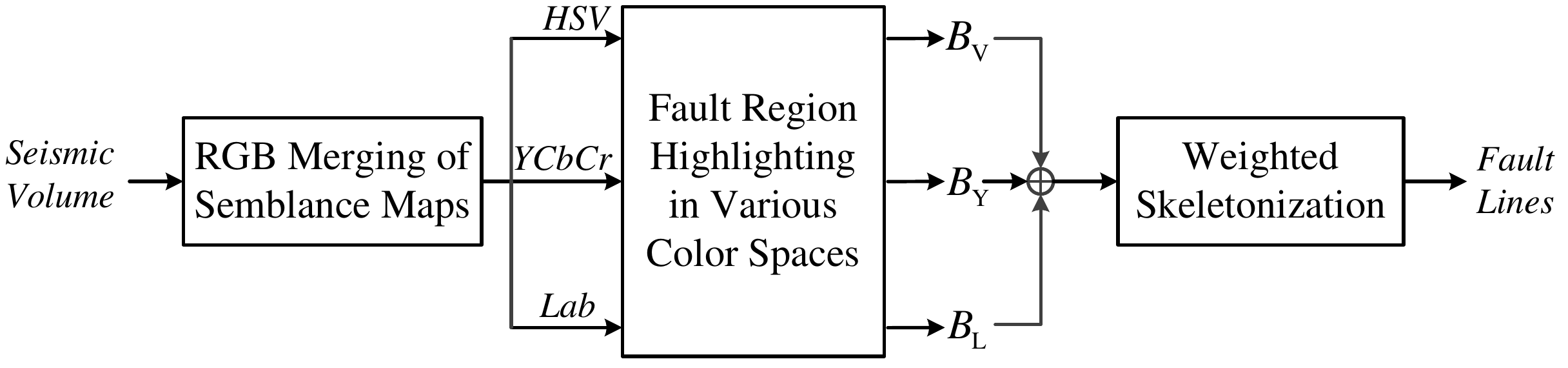}}
\end{minipage}
\caption{Block diagram of the proposed method}
\label{fig:diagram}
\end{figure}

The block diagram of the proposed method is shown in Fig.~\ref{fig:diagram}. We explain the main blocks of the proposed pipeline in the following subsections.

\subsection{RGB Blending of Semblance Maps}
\label{ssec:semb}
\begin{figure}[t]
\begin{minipage}[b]{.48\linewidth}
  \centering
  \centerline{\includegraphics[height=2cm]{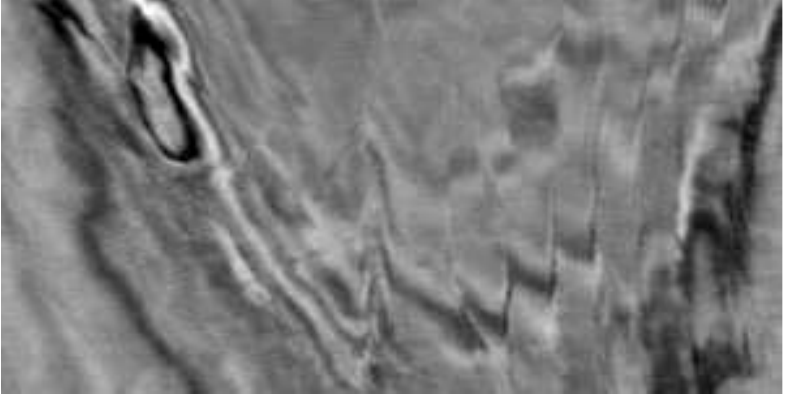}}
  \centerline{\small{(a) Time section $S_{t_0}$}}\medskip
\end{minipage}
\hspace{0.02in}
\begin{minipage}[b]{.48\linewidth}
  \centering
  \centerline{\includegraphics[height=2cm]{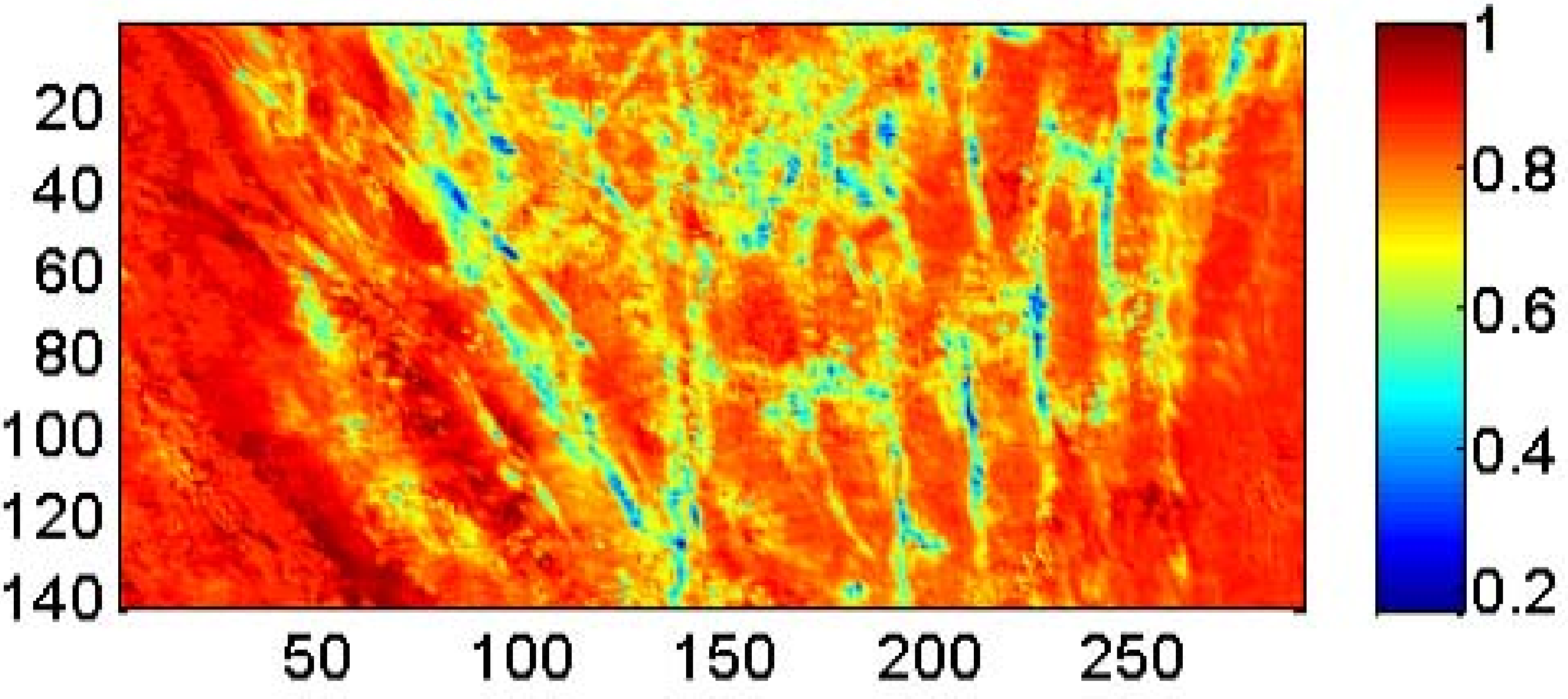}}
  \centerline{\small{(b) Semblance map $D_{t_0}$}}\medskip
\end{minipage}
\caption{The time section at $t_0$ and its semblance map}
\label{fig:section}
\end{figure}

To characterize fault regions in time sections, we utilize the most prominent feature of faults, discontinuities in horizons, which can be measured using the semblance attribute proposed by Marfurt et al.~\cite{marfurt1999coherency}. Semblance outperforms other seismic attributes in identifying the existence of faults by examining local dip information with a neighboring average. Fig.~\ref{fig:section} illustrates the time section at $t_0$, denoted $S_{t_0}$ and its corresponding semblance map $D_{t_0}$. As shown in Fig.~\ref{fig:section}(b), the red regions with high semblance values belong to horizons. In contrast, the green and blue regions with smaller semblance values represent likely fault regions.

Based on the semblance attribute, the algorithm proposed by Zhang et al.~\cite{zhang2013fault} extracts fault lines in a single time section. However, to obtain more accurate attribute maps, we need to use the highly correlated information of neighboring time sections. Time section $S_{t_0}$ has two neighboring sections, $S_{t_0-1}$ and $S_{t_0+1}$, the faults of which have shapes and structures similar to those of $S_{t_0}$, because of the consistent nature of geological structures. Since semblance maps were proposed based on geological structures, we select the semblance maps of three neighboring time sections, referred to as $D_{t_0-1}$, $D_{t_0}$, and $D_{t_0+1}$. Because of the high correlation between these neighboring semblance maps, we blend them as if they were red (\texttt{R}), green (\texttt{G}), and blue (\texttt{B}) channels of a single color image. The color-blended image with high contrast is shown in Fig.~\ref{fig:RGB}(a), in which black stripes correspond to likely fault regions. Compared to $D_{t_0}$, $C_{t_0}$ acts as a better indicator of likely fault regions. Thus, color-blended maps can increase the accuracy in fault detection for the interpreters.

\subsection{Fault Region Highlighting in Various Color Spaces}
\label{ssec:color}
\begin{figure}[t]
\begin{minipage}[b]{.48\linewidth}
  \centering
  \centerline{\includegraphics[width=4cm]{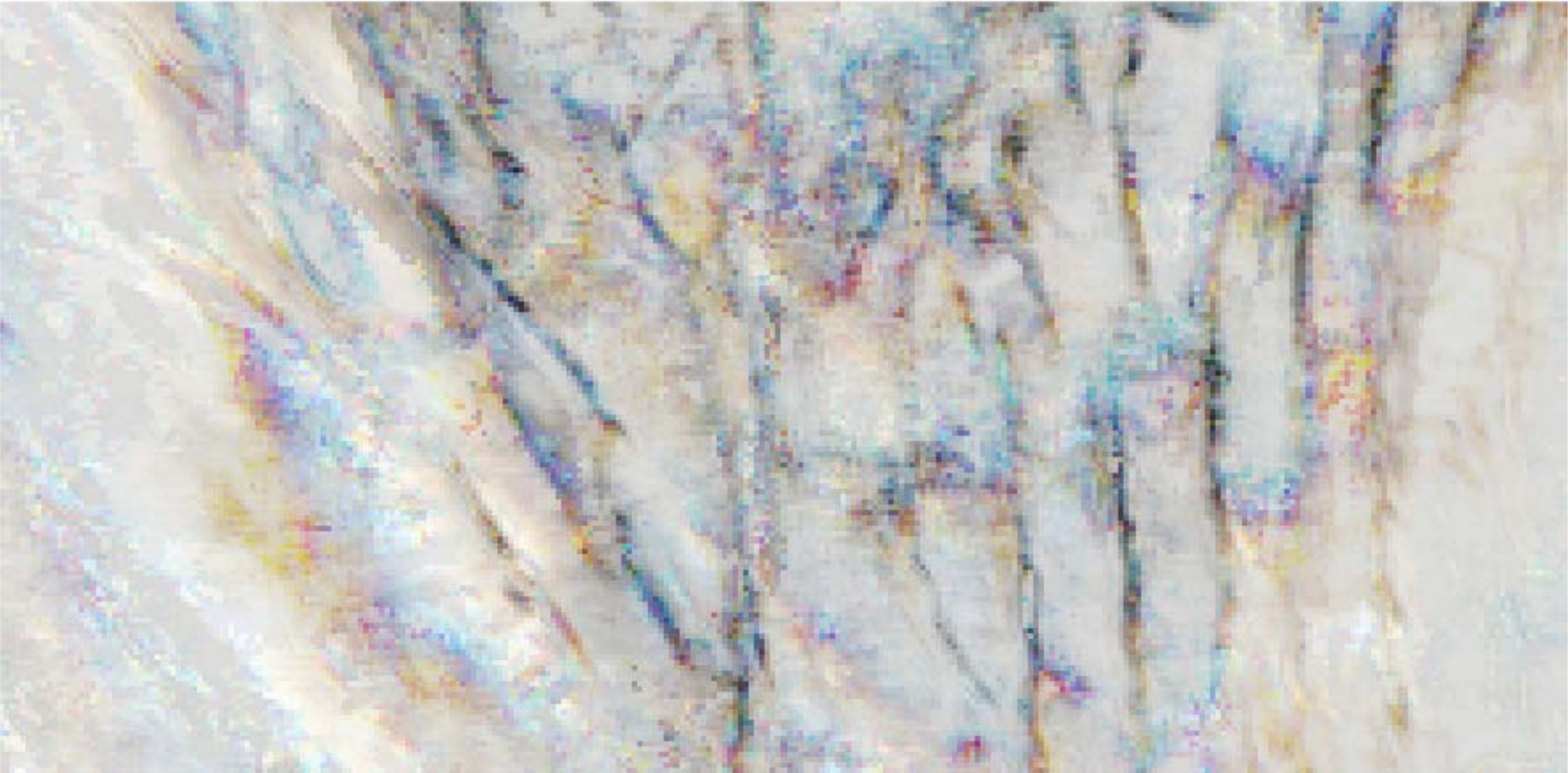}}
  \centerline{\footnotesize{(a) $C_{t_0}$ in \texttt{RGB} space}}\medskip
\end{minipage}
\hspace{0.02in}
\begin{minipage}[b]{.48\linewidth}
  \centering
  \centerline{\includegraphics[width=4cm]{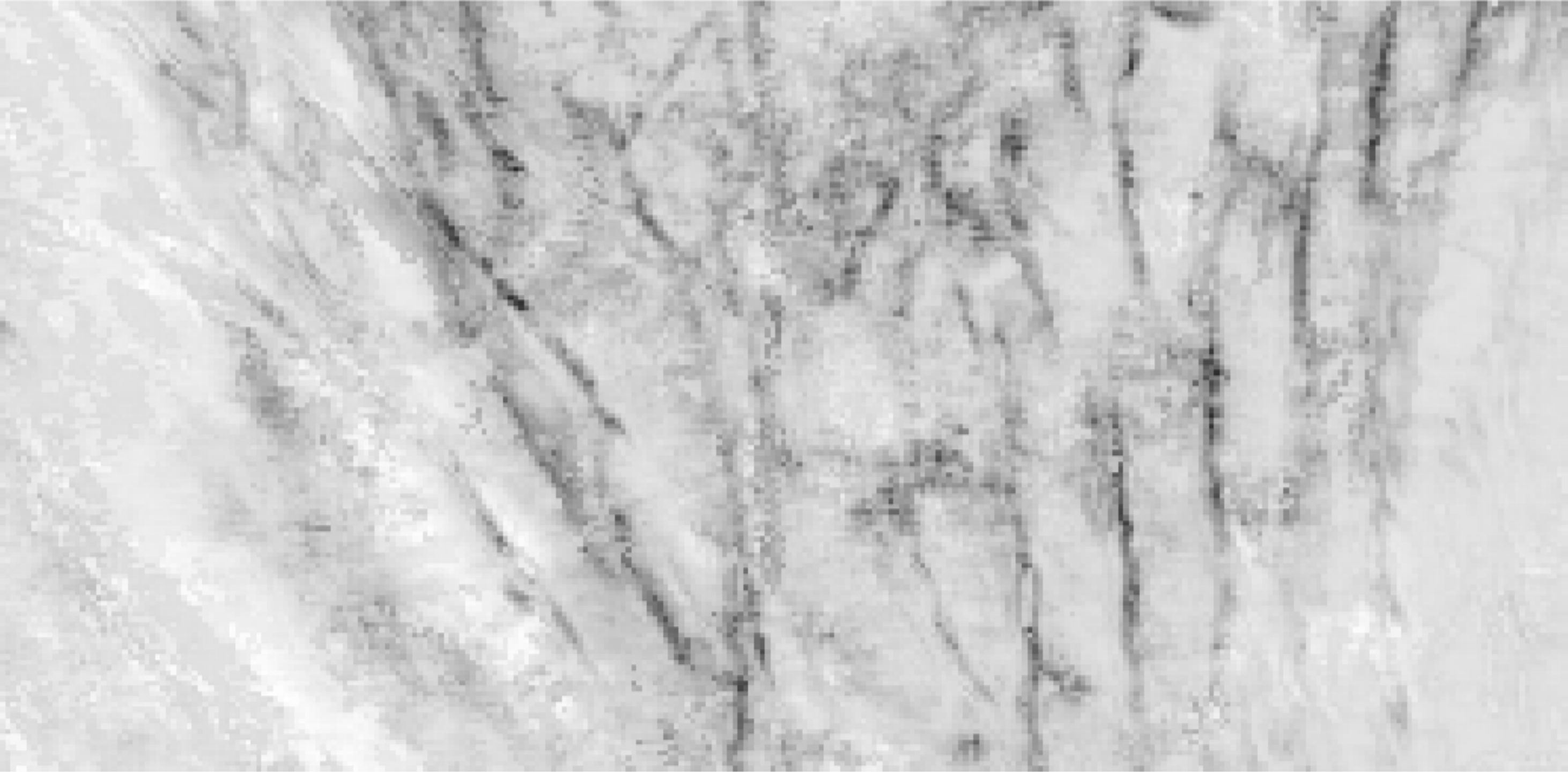}}
  \centerline{\footnotesize{(b) \texttt{L} channel in \texttt{Lab} model, $L_{t_0}$}}\medskip
\end{minipage}
\begin{minipage}[b]{.48\linewidth}
  \centering
  \centerline{\includegraphics[width=4cm]{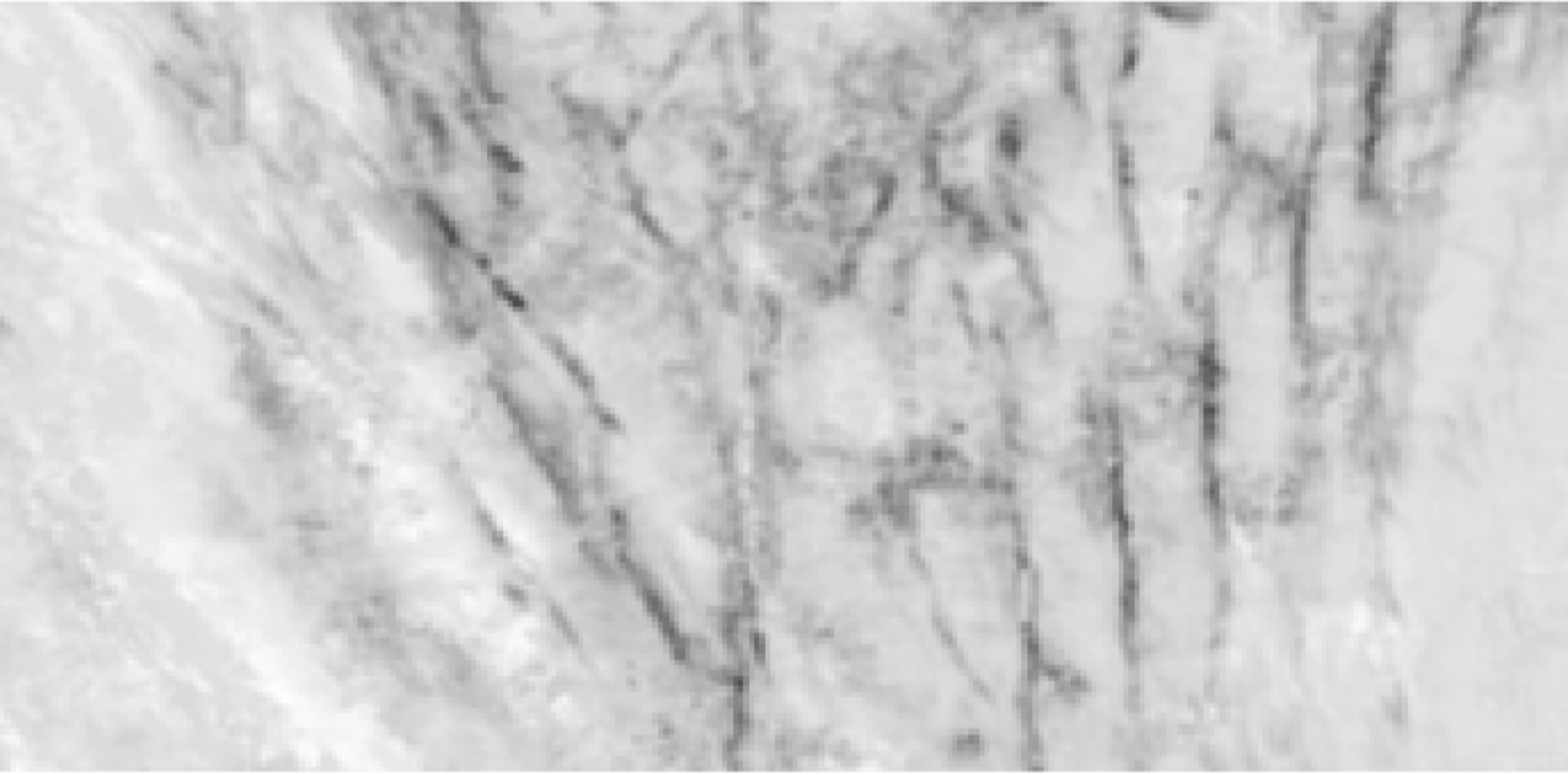}}
  \centerline{\footnotesize{(c) Smoothed \texttt{L} channel}}\medskip
\end{minipage}
\hspace{0.02in}
\begin{minipage}[b]{.48\linewidth}
  \centering
  \centerline{\includegraphics[width=4cm]{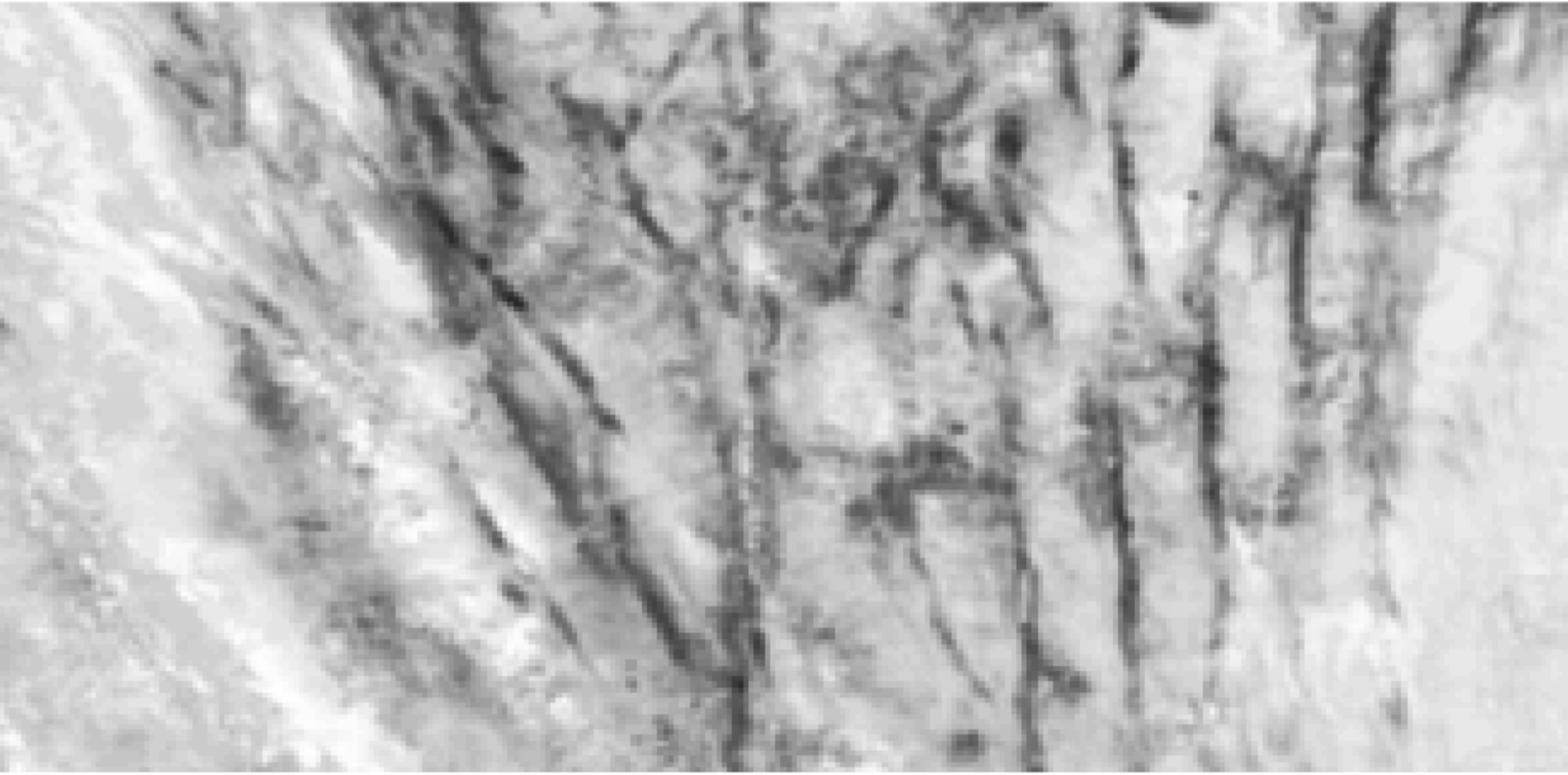}}
  \centerline{\footnotesize{(d) Smoothed \texttt{L} channel after CLAHE, $\hat{L}_{t_0}$}}\medskip
\end{minipage}
\caption{Color-blended image $C_{t_0}$ and the enhancement process of \texttt{L} channel}
\label{fig:RGB}
\end{figure}

Each channel in the \texttt{RGB} color space contains both chroma and luma information. In order to separate the color and structure-based components, we can use color space transformations. In the proposed pipeline, we transform the \texttt{RGB} images into \texttt{YCrCb}, \texttt{Lab}, and \texttt{HSV} spaces and utilize the luminance (\texttt{Y}), lightness (\texttt{L}), and value (\texttt{V}) channels from different color spaces. For the visualization of the main blocks in the pipeline, we use the lightness channel. However, the same steps are applicable to the luminance and value channels as well. An intensity map, denoted $L_{t_0}$, corresponding to the lightness channel is shown in Fig.~\ref{fig:RGB}(b), in which dark stripes indicate likely fault regions.

\begin{figure}[t]
\begin{minipage}[b]{.48\linewidth}
  \centering
  \centerline{\includegraphics[width=4cm]{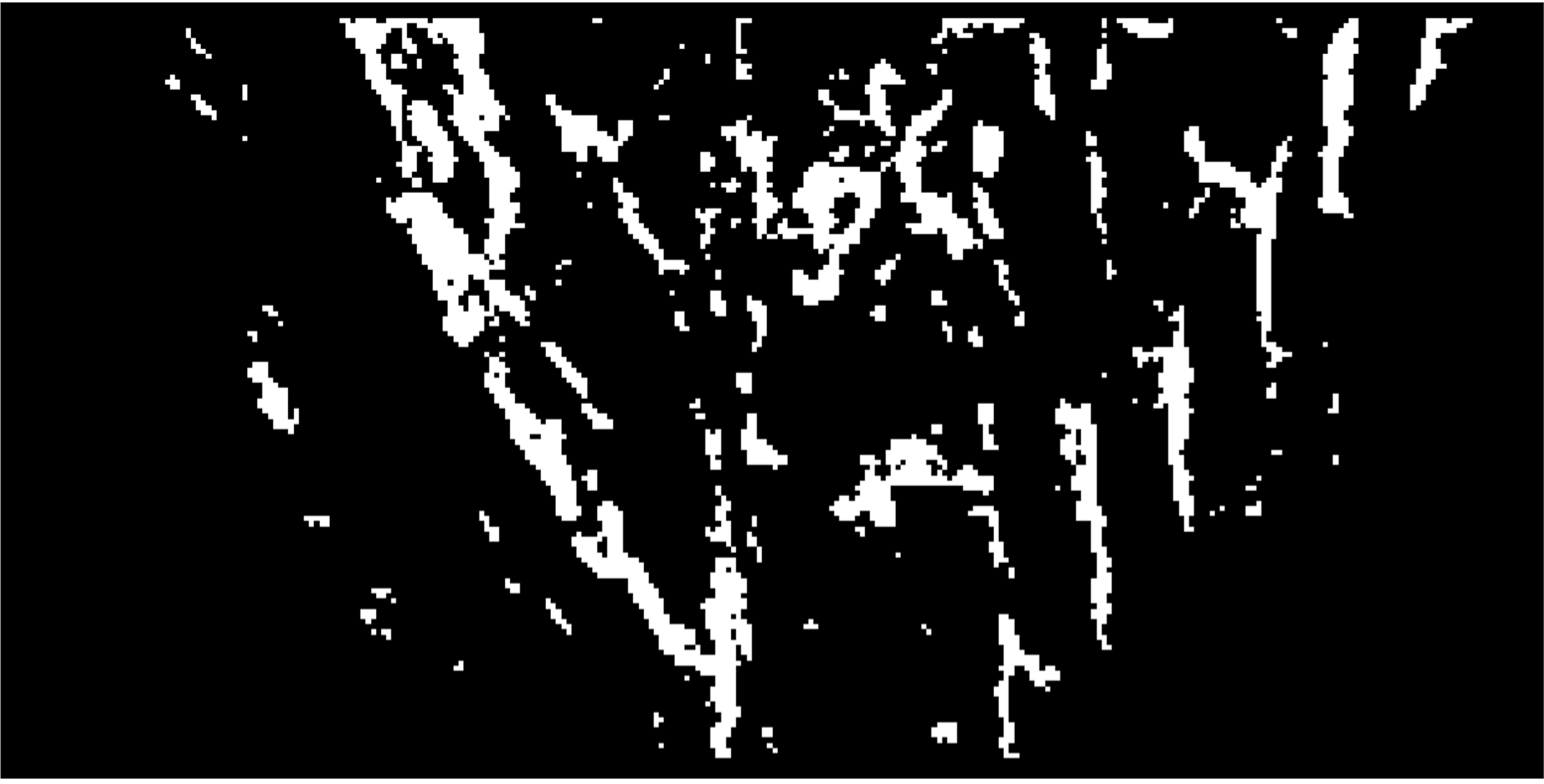}}
  \centerline{\footnotesize{(a) $B_{L,t_0}$}}\medskip
\end{minipage}
\hspace{0.02in}
\begin{minipage}[b]{.48\linewidth}
  \centering
  \centerline{\includegraphics[width=4cm]{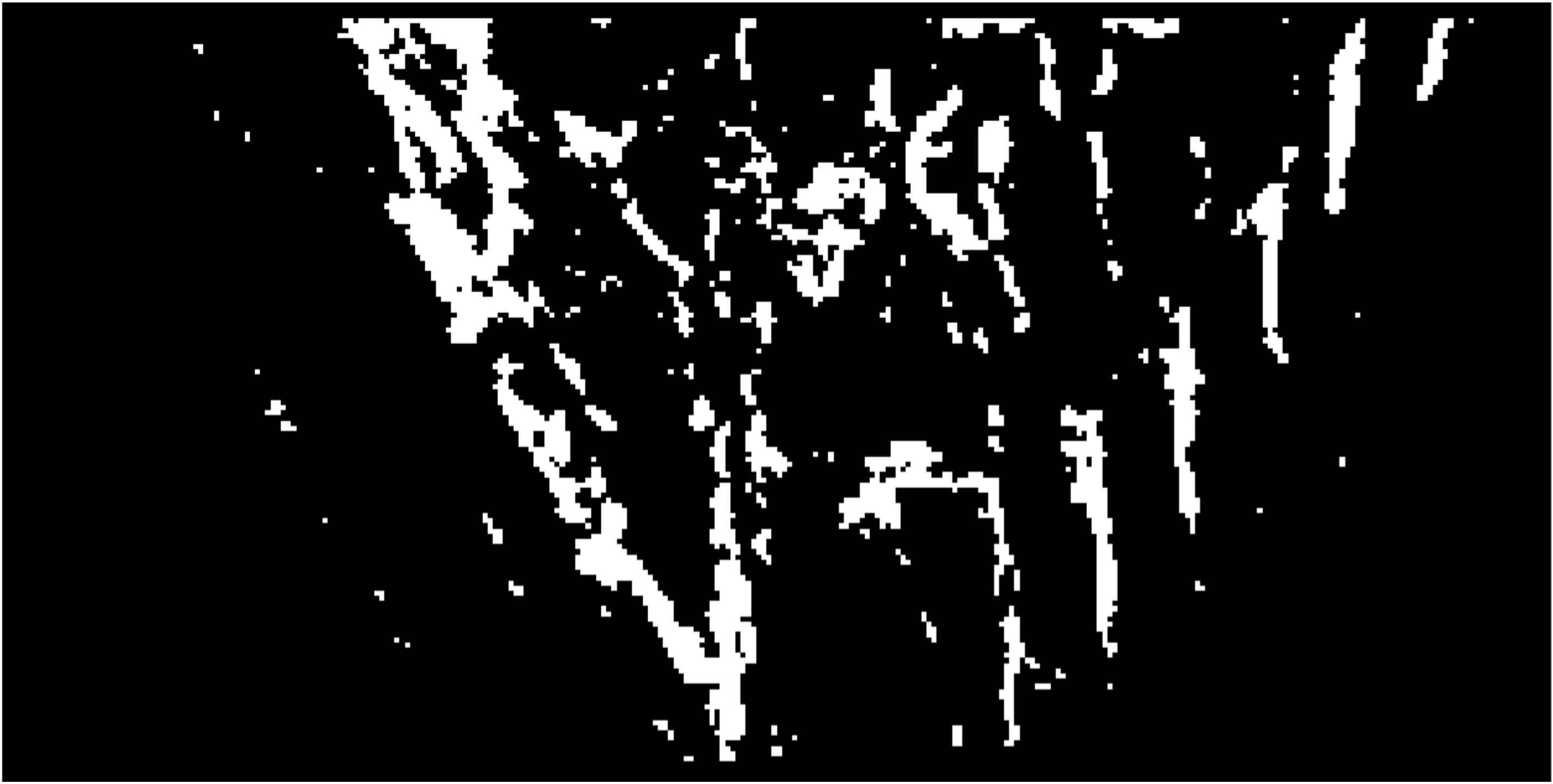}}
  \centerline{\footnotesize{(b) $B_{V,t_0}$}}\medskip
\end{minipage}
\begin{minipage}[b]{.48\linewidth}
  \centering
  \centerline{\includegraphics[width=4cm]{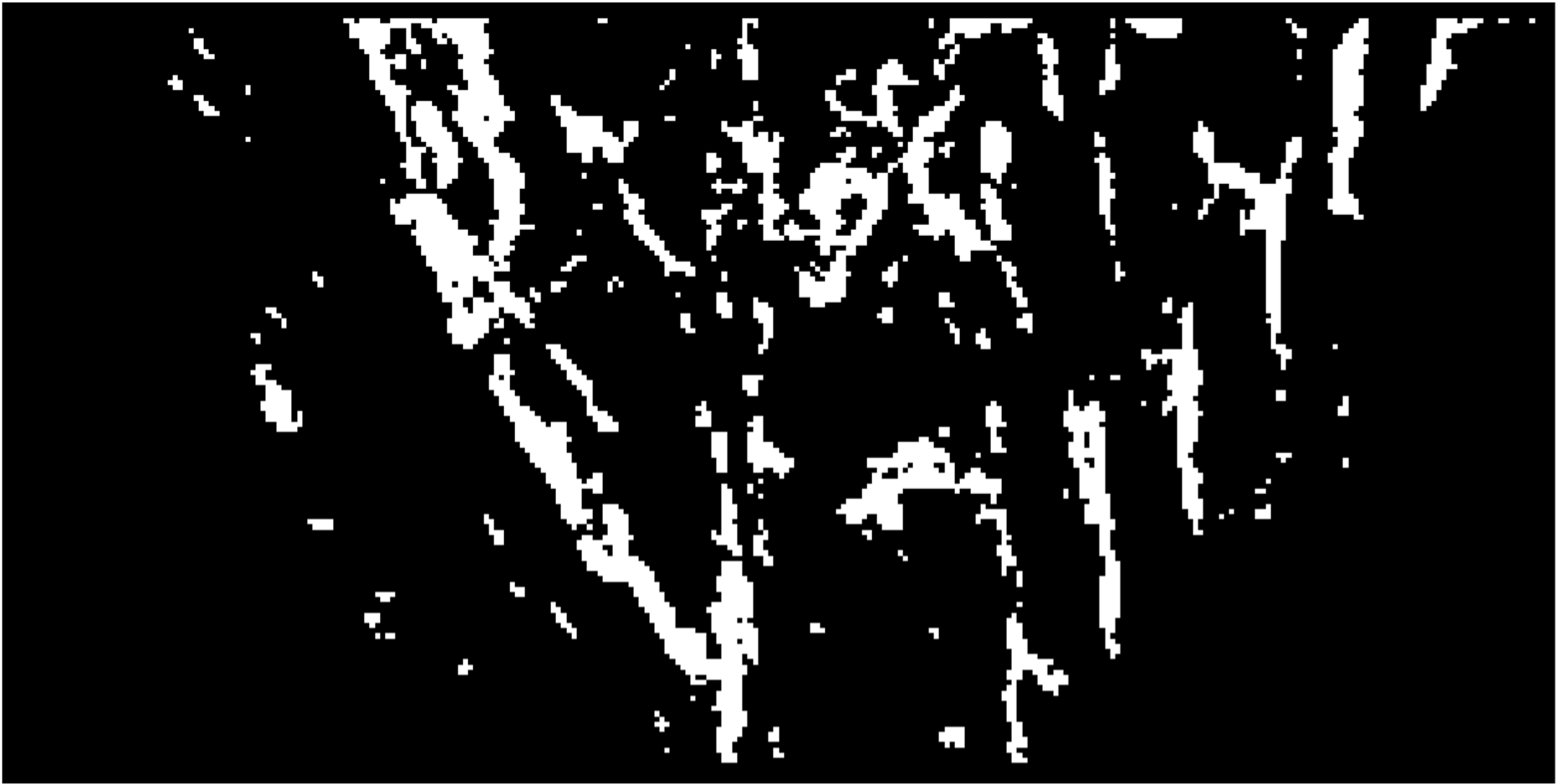}}
  \centerline{\footnotesize{(c) $B_{Y,t_0}$}}\medskip
\end{minipage}
\hspace{0.02in}
\begin{minipage}[b]{.48\linewidth}
  \centering
  \centerline{\includegraphics[width=4cm]{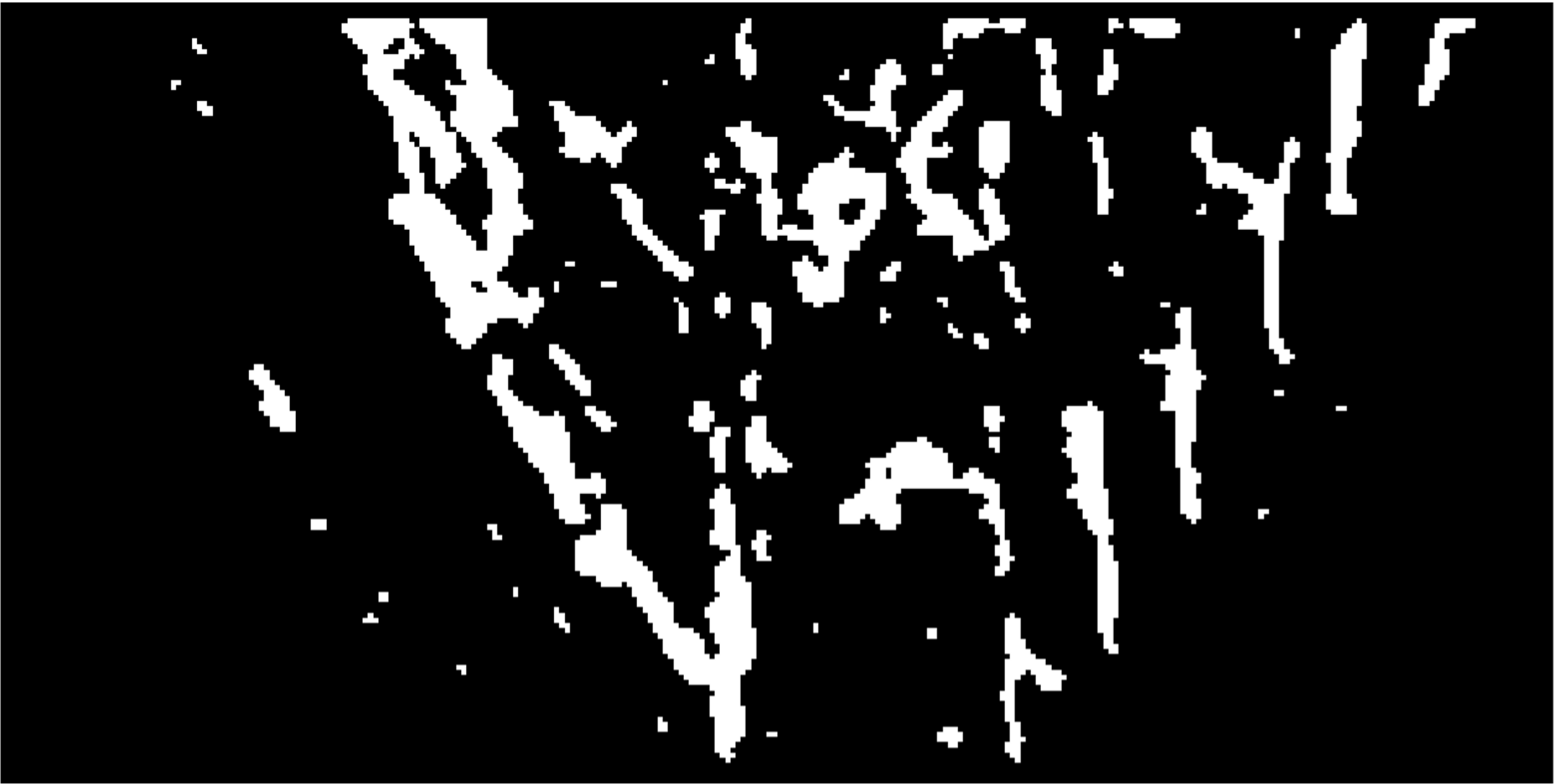}}
  \centerline{\footnotesize{(d) $B_{t_0}$}}\medskip
\end{minipage}
\caption{Highlighted fault regions in different channels ($B_{L,t_0}, B_{V,t_0}, \mbox{and }B_{Y,t_0}$) and the combined results in $B_{t_0}$.}
\label{fig:binary}
\end{figure}

To remove the noise around the likely fault regions, we smooth lightness channel $L_{t_0}$ using a Gaussian kernel with standard derivation $\sigma$ and size $r\times r$, and the smoothed lightness map is shown in Fig.~\ref{fig:RGB}(c). Furthermore, to enhance the contrast between faults and horizons, we utilize contrast limited adaptive histogram equalization (CLAHE)~\cite{zuiderveld1994contrast}. The main advantage of CLAHE, compared to other histogram equalization methods, comes from the contrast threshold that prevents the majorities from biasing the equalization. In the case of seismic maps, CLAHE eliminates the contrast enhancement of horizon regions and boosts the contrast for likely fault regions. The enhanced lightness channel, denoted as $\hat{L}_{t_0}$, is shown in Fig.~\ref{fig:RGB}(d), in which the likely fault regions become more distinguishable. To highlight these candidates of fault regions, we apply a threshold, denoted $T_L$, on enhanced lightness map $\hat{L}_{t_0}$ and obtain a binary map $B_{L,t_0}$ as shown in Fig.~\ref{fig:binary}(a). The thresholding process is formulated in Eq.~(\ref{equ:thres}).
\begin{equation}
\centering
\label{equ:thres}
B_{L,t_0} (x, y) =
\left\{
\begin{aligned}
&1,\quad \mbox{if } \hat{L}_{t_0} (x,y)<T_L\\
&0,\quad \mbox{otherwise}
\end{aligned}
\right.
,
\end{equation}
where $x$ and $y$ represent the inline and crossline directions, respectively. The same procedure is applied on the luminance and value channels and the corresponding binary maps are denoted as $B_{Y,t_0}$ and $B_{V,t_0}$, shown in Fig.~\ref{fig:binary}(b) and Fig.~\ref{fig:binary}(c), respectively. Since all of these binary images contain similar fault structures, the combination of
these images can lead to more accurate fault region candidates. Although adding is a straightforward way to combine these binary images, it may amplify noise around fault regions. Therefore, we propose combining these images under geological constraints as follows:
\begin{equation}
\label{equ:combine1}
\Scale[0.75]{
B_{t_0}(x,y) =
\left\{
\begin{aligned}
&1,\mbox{ if }\sum_{i=L, Y, V}B_{i, t_0}(x,y)\geq 2,\mbox{ and }D_{t_0}(x,y)\leq T_C \\
&1,\mbox{ if }\sum_{i=L, Y, V}B_{i, t_0}(x,y)= 1\\
&0,\mbox{ otherwise}
\end{aligned}
\right.}
,
\end{equation}
where $T_C$ is a threshold to filter out noisy points with greater semblance values. Eq.~(\ref{equ:combine1}) indicates that a pixel belongs to fault regions if its semblance value is less than $T_C$ and it appears in at least two channels. Moreover, to ensure the connectivity of fault regions, we also consider pixels detected in only one channel as fault regions. Under the constraints in Eq.~(\ref{equ:combine1}), we obtain the combined binary image $B_{t_0}(x,y)$, shown in Fig.~\ref{fig:binary}(d).

\subsection{Weighted Skeletonization}
\label{ssec:skeleton}

To label one-pixel-width fault lines from highlighted fault regions, we need to apply skeletonization, a thinning process extracting topological skeletons of shapes, on $B_{t_0}$. The skeleton of a 2D shape is composed of the locus of the centers of all maximum inscribed disks, which can not be covered by any other inscribed disks and have at least two tangential points with the boundaries of the shape. In this paper, we propose a weighted skeletonization method to delineate fault lines more accurately by involving geological constraints.  Our method is based on the Voronoi diagram, a powerful tool on implementing skeletonization~\cite{brandt1992continuous}.
However, the skeletons only extracted from the Voronoi diagram are not accurate enough to represent the structure of faults because of the undesired branches.

To remove these undesired branches, we define weight $W_{t_0}(x,y)$ at every point of the initially extracted skeletons as the multiplication of two indices explained in Eq.~(\ref{equ:indices}):
\begin{equation}
\label{equ:indices}
\centering
W_{t_0}(x,y) = K_{t_0}(x,y)\times G_{t_0}(x,y),
\end{equation}
where $K_{t_0}(x,y)$ and $G_{t_0}(x,y)$ represent the dimensional and geological weights, respectively. In the maximum inscribed disk of $(x,y)$, $K_{t_0}(x,y)$ is defined as the length of the longest arc between two neighboring tangential points~\cite{brandt1992continuous}. By indicating the dimension of disks, $K_{t_0}(x,y)$ plays an important role in distinguishing undesired branches near vertices.
%\begin{figure}[t]
%%
%\begin{minipage}[b]{1.0\linewidth}
%  \centering
%  \centerline{\includegraphics[width=6cm]{Figs/skeleton}}
%\end{minipage}
%\caption{Skeletonization based on Voronoi diagram}
%\label{fig:skeleton}
%%
%\end{figure}
\begin{figure}[t]
\begin{minipage}[b]{.48\linewidth}
  \centering
  \centerline{\includegraphics[width=4cm]{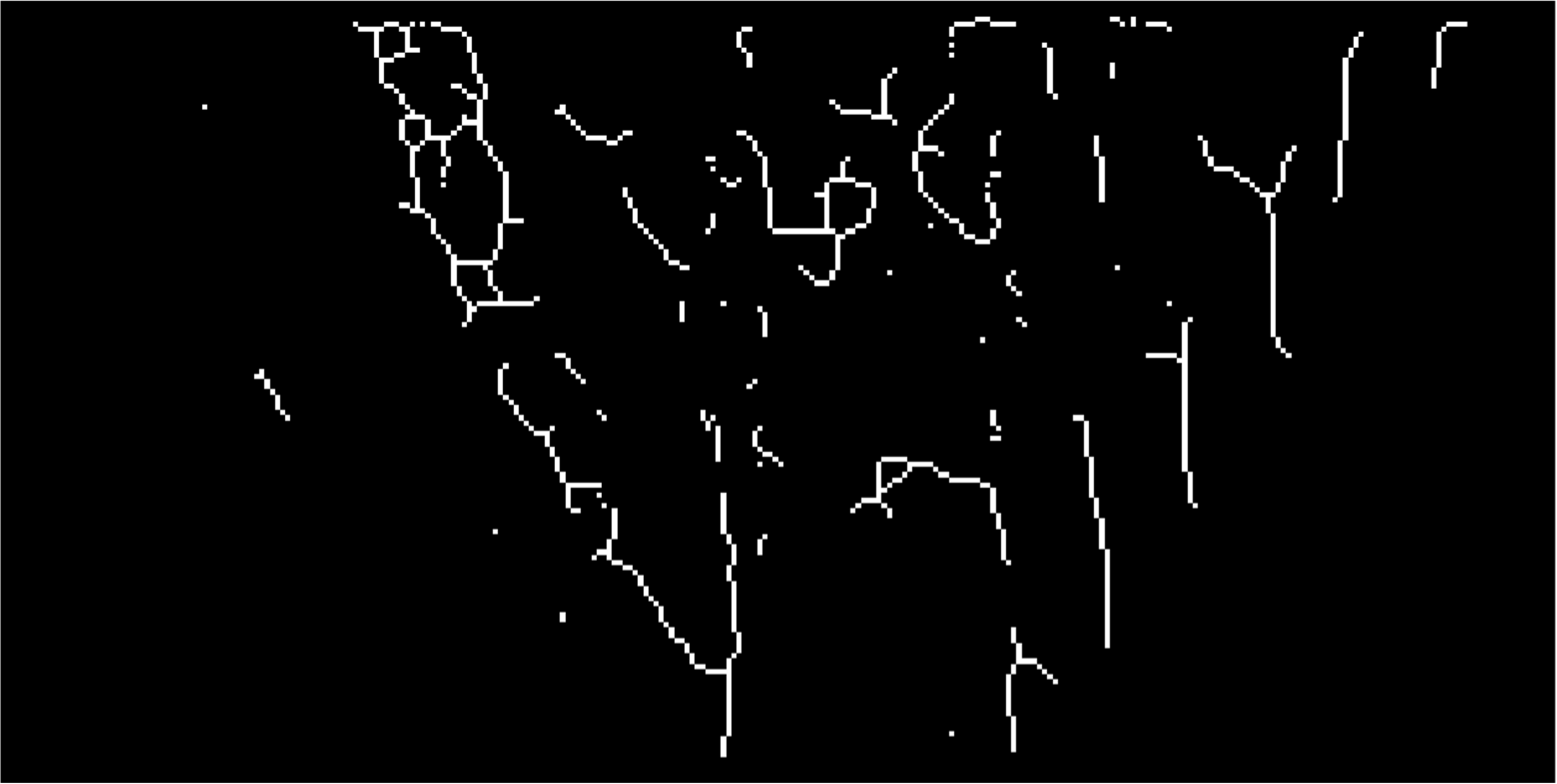}}
  \centerline{\footnotesize{(a) Skeletonization weighted by $W_{t_0}$}}\medskip
\end{minipage}
\hspace{0.05in}
\begin{minipage}[b]{.48\linewidth}
  \centering
  \centerline{\includegraphics[width=4cm]{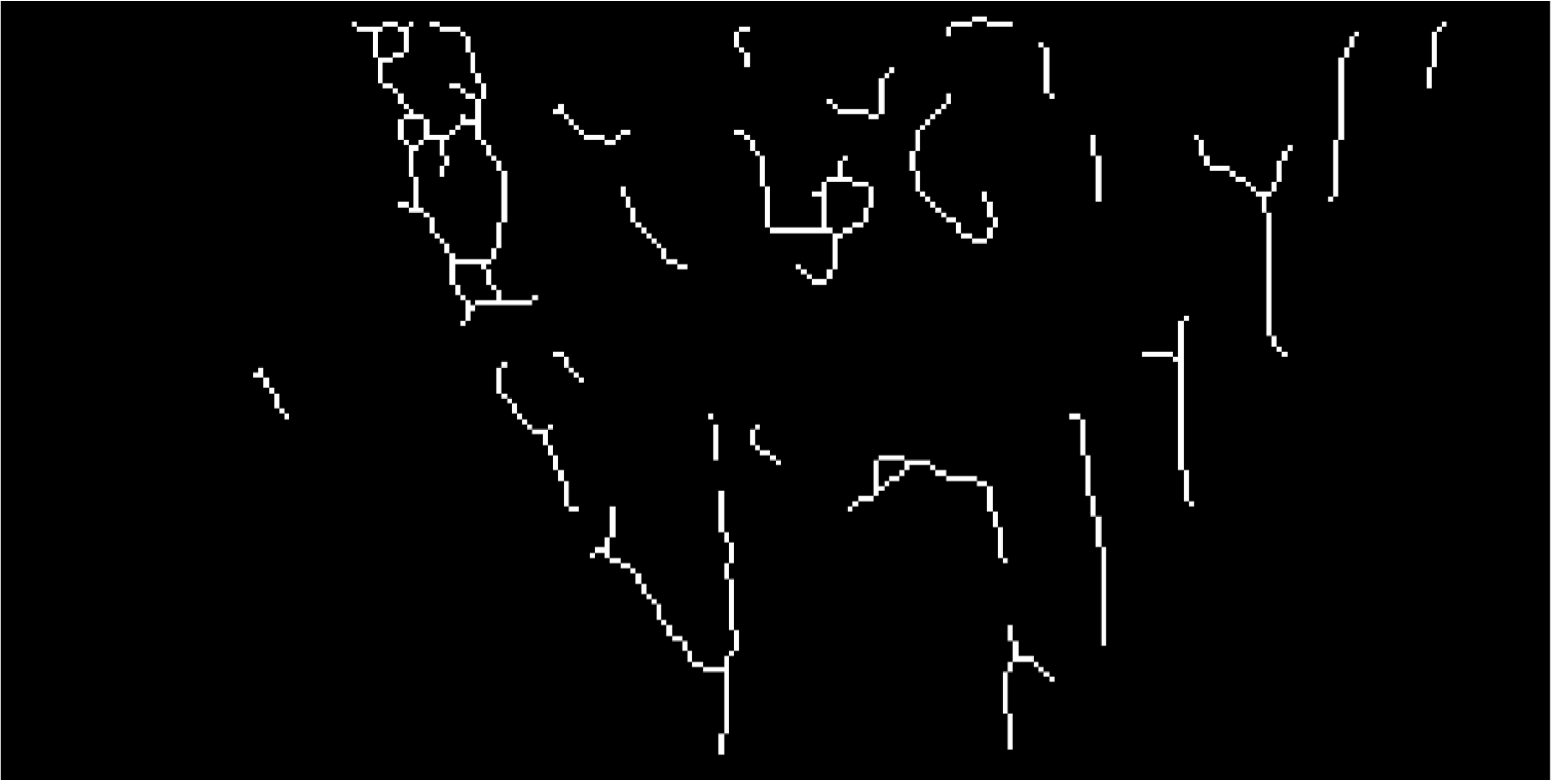}}
  \centerline{\footnotesize{(b) Extracted fault lines}}\medskip
\end{minipage}
\caption{Fault extraction based on the weighted skeletonization}
\label{fig:eg}
\end{figure}

Because of the intricate shapes of highlighted fault regions, we can not easily prune all noisy branches based only on the dimensional index. Therefore, we propose the geological index based on the semblance map to remove branches located around the fault regions with high semblance values, which correspond to low discontinuities. Since fault regions highlighted in $B_{t_0}$ are the combination of different color channels, we propose a discontinuity map ($\hat{D}_{t_0}$) that incorporates neighboring semblance information, which is calculated as follows:
\begin{equation}
\label{equ:max}
\centering
\hat{D}_{t_0}(x,y) = \underset{i\in[-1,0,1]}{\max}\left|\log(D_{t_0+i}(x,y))\right|,
\end{equation}
where $\hat{D}_{t_0}(x,y)$ corresponds to the largest discontinuity value in three neighboring time sections.
Then, to remove noise and enhance discontinuities, we smooth $\hat{D}_{t_0}(x,y)$ by averaging it in its square neighborhood weighted by the power of the intensity of seismic signals, as shown in Eq.~(\ref{equ:weighted}):
\begin{equation}
\label{equ:weighted}
\centering
\Scale[1.2]{
G_{t_0}(x,y) = \frac{\sum\limits_{i,j=-r_s}^{r_s}\hat{D}_{t_0}(x+i,y+j)\cdot S_{t_0}^2(x+i,y+j)}{\sum\limits_{i,j=-r_s}^{r_s}S_{t_0}^2(x+i,y+j)}},
\end{equation}
where $G_{t_0}(x,y)$ corresponds to the obtained geological index of point $(x,y)$ and $r_s$ determines the size of the square neighborhood. Using this index, a point with larger weight $W_{t_0}(x,y)$ corresponds to a larger inscribed disk and greater discontinuity value and has a higher probability of being located on a fault. By applying a global threshold $T_W$ on the weights of the initially extracted skeletons, we obtain the binary image $I_{t_0}$ containing the pruned skeletons as follows:
\begin{equation}
\label{equ:globalThres}
I_{t_0}(x,y)=
\left\{
\begin{aligned}
&1,\quad\mbox{if } W_{t_0}(x,y)\geq T_W\mbox{,}\\
&0,\quad\mbox{otherwise,}
\end{aligned}
\right.
\end{equation}
where $T_W$ is set empirically by interpreters. $I_{t_0}$ in Fig.~\ref{fig:eg}(a) illustrates the extracted fault lines with most noisy branches removed.
After removing the isolated line segments and short branches, we obtain the smoothed delineation of faults in time section $t_0$, as Fig.~\ref{fig:eg}(b) shows.

\section{Experimental Results}
\label{sec:results}
In this paper, we applied the proposed algorithm on time sections of the 3D seismic data set acquired from the Netherlands offshore F3 block in the North Sea~\cite{f3opendtect}. The tested 3D seismic volume, a local region extracted from F3 block, contains distinguishable fault structures and has the dimension ranging from \#199 to \#349 in the inline direction, from \#300 to \#599 in the crossline direction, and from 1396ms to 1848ms in the time direction with the step of 4ms.

To illustrate the performance of the proposed algorithm, we take the time section at $t_0=1604\mbox{ms}$ as an example.
As Fig.~\ref{fig:section}(a) shows, discontinuous regions in the time section $S_{t_0}$ indicate the existence of faults.
The semblance map $D_{t_0}$ in Fig.~\ref{fig:section}(b) illustrates contrast between likely fault regions and horizons. To involve more structural information of faults, we blend three neighboring semblance maps into a color image, $C_{t_0}$, in \texttt{RGB} model as Fig.~\ref{fig:RGB}(a) shows. Then, we transfer $C_{t_0}$ from the \texttt{RGB} model to the \texttt{Lab}, \texttt{YCbCr}, and \texttt{HSV} models, all of which contain separated intensity component, referred to as the \texttt{L}, \texttt{Y}, and \texttt{V} channels, respectively. \texttt{L} channel, as an example of the intensity component, is shown in Fig.~\ref{fig:RGB}(b). To remove the noise around likely fault regions in the \texttt{L} channel, we adopt a $2\times2$ Gaussian filter with $\sigma=10$. In addition, by applying the CLAHE on the smoothed \texttt{L} channel in Fig.~\ref{fig:RGB}(c), we obtain the enhanced likely fault regions in Fig.~\ref{fig:RGB}(d). Furthermore, we set threshold $T_L=0.55$ on $\hat{L}_{t_0}$ to highlight fault regions in binary image $B_{L,t_0}$. Similarly, we apply smoothing, CLAHE, and thresholding on the \texttt{Y} and \texttt{V} channels and obtain two more binary images $B_{Y,t_0}$ and $B_{V,t_0}$. All parameters involved in the CLAHE are set empirically by interpreters and remain unchanged for the other two channels. However, we need to tweak the highlighting thresholds in different channels because of the ranges of different color spaces. The combination of these binary images leads to $B_{t_0}$ in Fig.~\ref{fig:binary}(d), which contains fault regions with the highest accuracy. Finally, we apply the weighted skeletonization on $B_{t_0}$ and extract fault lines shown in Fig.~\ref{fig:eg}(a). In Fig.~\ref{fig:eg}(b), we further remove isolated line segments and short branches to smoothen the extracted results.

\begin{figure}[t]
\begin{minipage}[b]{.48\linewidth}
  \centering
  \centerline{\includegraphics[height=2.2cm]{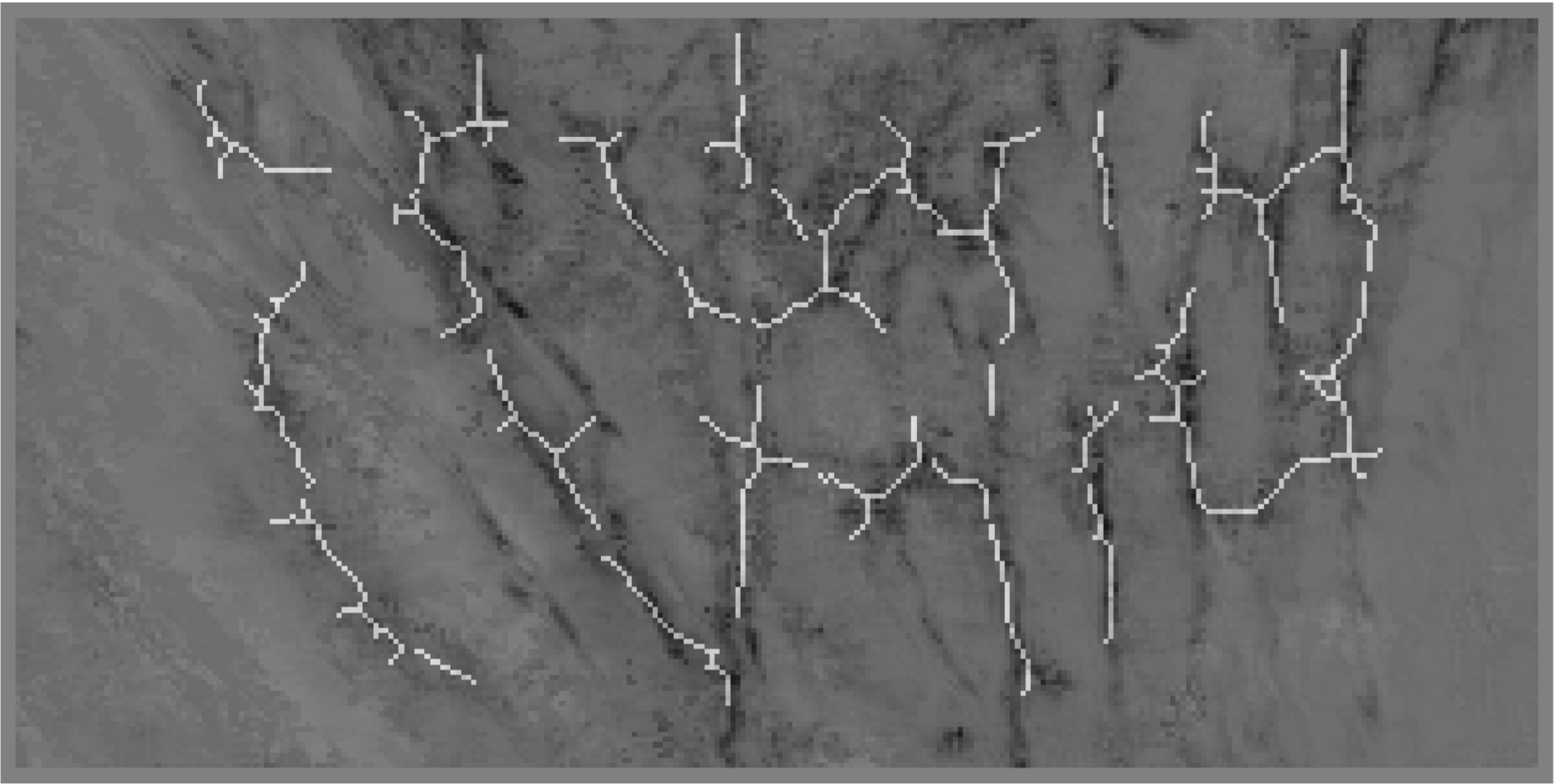}}
  \centerline{\small{(a) Results of the proposed method}}\medskip
\end{minipage}
\hspace{0.05in}
\begin{minipage}[b]{.48\linewidth}
  \centering
  \centerline{\includegraphics[height=2.2cm]{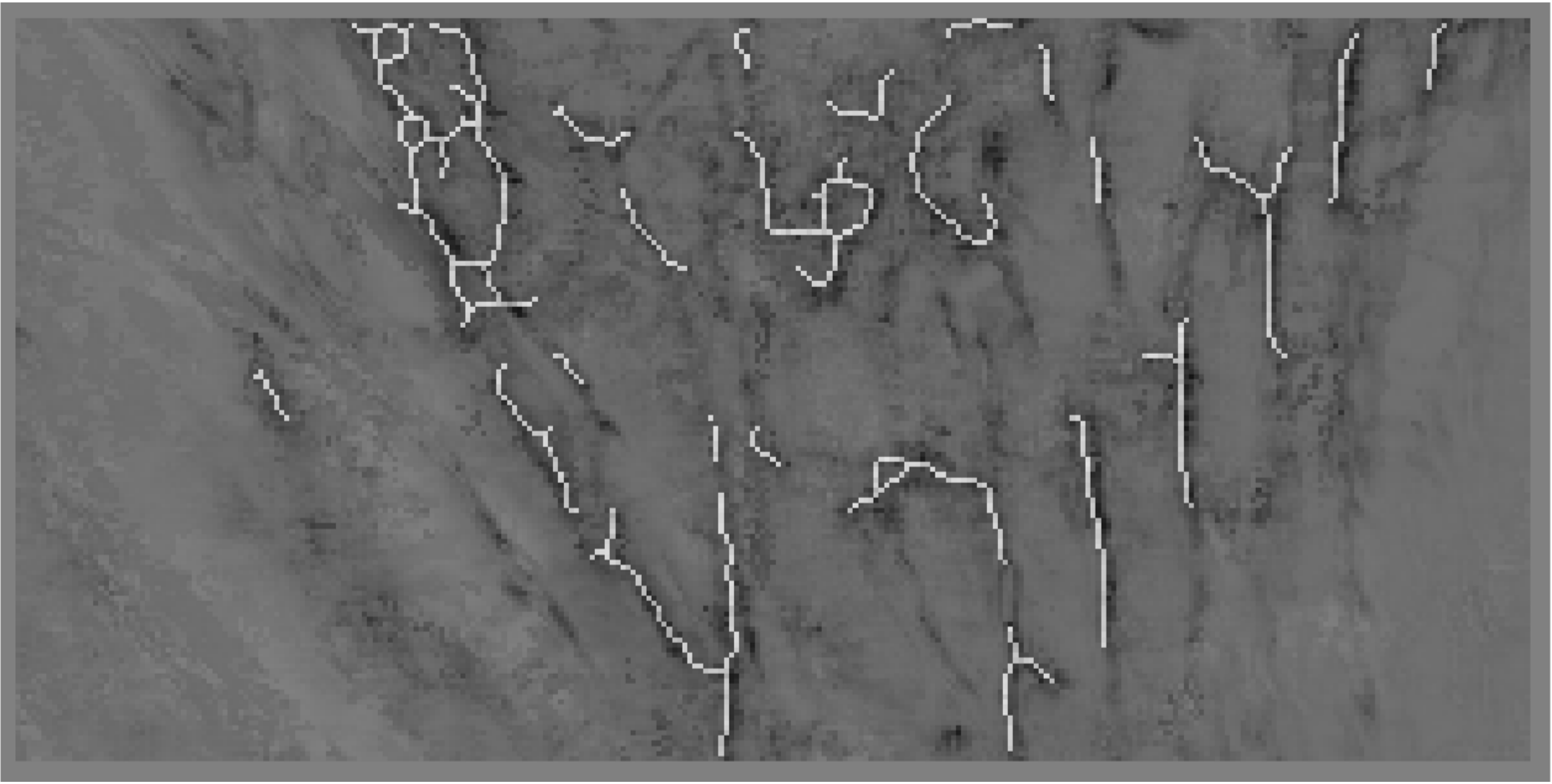}}
  \centerline{\small{(b) Results of the method in~\cite{zhang2013fault}}}\medskip
\end{minipage}
\caption{The comparison of different fault detection methods in the time section at 1604ms}
\label{fig:cmpr1}
\end{figure}

To clearly visualize and compare the performance of different methods, as Figs.~\ref{fig:cmpr1} shows, we merge the extracted fault lines into the corresponding semblance map where light regions indicate horizons and dark regions imply faults.
We recognize that the fault lines extracted by the proposed method in Figs.~\ref{fig:cmpr1}(a) almost cover all possible fault regions and have smooth outlines. In contrast, the method proposed in~\cite{zhang2013fault}
mistakenly detects fault lines in horizons and generates noisy branches.
Though Zhang's method is very robust and requires limited human intervention, the proposed method leads to a higher detection accuracy by involving color representations and geological constraints.
To quantitatively measure the difference between the detected results and the ground truth, we define the distance between two points $(x_1,y_1)$ and $(x_2,y_2)$ as follows:
$dist=\min\left(\left|x_1-x_2\right|,\left|y_1-y_2\right|\right)$. We select several time sections and calculate the corresponding average distances in Table~\ref{tab:tab1}. The first and third column represent the average distances based on the proposed method and the method in~\cite{zhang2013fault}, respectively. In addition, the second column corresponds to the distances calculated based on the proposed method without involving color blending and color transformations. This difference is primarily accredited to the usage of color representations and the removal of noisy branches using the semblance index. Furthermore, the highly parallel structure of the proposed algorithm ensures the real-time implementation in semi-automatic seismic interpretation, but this feature will be considered in a future work focusing on the efficiency.
%\begin{figure}[t]
%%
%\begin{minipage}[b]{.48\linewidth}
%  \centering
%  \centerline{\includegraphics[height=2.2cm]{Figs/cmpr1}}
%  \centerline{\footnotesize{(a) Results of the proposed method}}\medskip
%\end{minipage}
%\hspace{0.05in}
%\begin{minipage}[b]{.48\linewidth}
%  \centering
%  \centerline{\includegraphics[height=2.2cm]{Figs/cmpr1}}
%  \centerline{\footnotesize{(b) Results without color representations}}\medskip
%\end{minipage}
%\begin{minipage}[b]{.48\linewidth}
%  \centering
%  \centerline{\includegraphics[height=2.2cm]{Figs/cmpr2}}
%  \centerline{\footnotesize{(c) Results of the method in~\cite{zhang2013fault}}}\medskip
%\end{minipage}
%\hspace{0.05in}
%\begin{minipage}[b]{.48\linewidth}
%  \centering
%  \centerline{\includegraphics[height=2.2cm]{Figs/cmprTruth}}
%  \centerline{\footnotesize{(d) The ground truth}}\medskip
%\end{minipage}
%\caption{The comparison of different fault detection methods in the time section at 1604ms}
%\label{fig:cmpr1}
%\end{figure}
%\begin{figure}[t]
%%
%\begin{minipage}[b]{.48\linewidth}
%  \centering
%  \centerline{\includegraphics[height=1.8cm]{Figs/cmpr3}}
%  \centerline{\small{(a) Results of the proposed method}}\medskip
%\end{minipage}
%\hspace{0.05in}
%\begin{minipage}[b]{.48\linewidth}
%  \centering
%  \centerline{\includegraphics[height=1.8cm]{Figs/cmpr4}}
%  \centerline{\small{(b) Results of the method in~\cite{zhang2013fault}}}\medskip
%\end{minipage}
%\caption{The comparison of different fault detection methods in the time section at 1576ms}
%\label{fig:cmpr2}
%\end{figure}

\vspace{-0.1in}
\begin{table}[!htbp]
\caption{Objective assessment of different methods}
\centering
\begin{threeparttable}
\begin{tabular}{c c c c}
\toprule
Time Sections & Proposed$^1$ & Proposed$^2$ & Zhang et al.~\cite{zhang2013fault}\\ [0.5ex]
\midrule
1576ms & \textbf{0.8682} & 1.2655 & 1.5064 \\
1604ms & \textbf{0.9236} & 1.1838 & 1.8217 \\
1624ms & \textbf{0.9305} & 0.9987 & 1.2582 \\ [0.5ex]
\bottomrule
\end{tabular}
\end{threeparttable}
\begin{tablenotes}[para,flushleft]
Note: $1$: proposed method with color blending and color transformations involved, $2$: proposed method without involving color representations.
\end{tablenotes}
\label{tab:tab1}
\end{table}
\vspace{-0.2in}

\section{Conclusion}
\label{sec:conclusion}

In this paper, we combined color blending and color transformations to semi-automatically detect faults in time sections . We first blended the semblance maps of neighboring time sections to synthesize a color image in the \texttt{RGB} model. By transforming the \texttt{RGB} image to \texttt{Lab}, \texttt{YCbCr}, and \texttt{HSV}, we obtained the separated intensity components, which contain important structural information related to faults. After the smoothing, enhancement, and thresholding, we highlighted the likely fault regions in  binary maps. Finally, we proposed the weighted skeletonization to extract one-pixel-width fault lines. Experimental results show that the proposed method improves the accuracy of the fault detection by limiting the average distance between detected fault lines and the ground truth into one pixel. Ongoing work focuses on the application of color blending and color transformations for semi-automatically detecting other seismic structures.

\end{document}